\documentclass[pre,aps,pacs,twocolumn]{revtex4}
\usepackage{epsfig}
\usepackage{bm}
\usepackage{color}

\newcommand{\B}[1]{{\bm{#1}}}

\usepackage[latin1]{inputenc}
\newcommand{\beq}{\begin{equation}}
\newcommand{\eeq}{\end{equation}}
\newcommand{\bea}{\begin{eqnarray}}
\newcommand{\eea}{\end{eqnarray}}

\begin{document}
\title{Ageing and Relaxation in Glass Forming Systems}
\author{Valery Ilyin,  Itamar Procaccia, Ido Regev and Nurith Schupper }
\affiliation{Department of Chemical Physics, The Weizmann
Institute of Science, Rehovot 76100, Israel }
\begin{abstract}
We propose that there exists a generic class of glass forming systems that have competing states (of crystalline order or not) which are locally close in energy to  the ground state (which is typically unique). Upon cooling, such systems exhibit patches (or clusters) of these competing states which become locally stable in the sense of having a relatively high local shear modulus. It is in between these clusters where ageing, relaxation and plasticity under strain can take place. We demonstrate explicitly that relaxation events that lead to ageing occur where the local shear modulus is low (even negative), and result in an increase in the size of local patches of relative order. We examine the ageing events closely from two points of view. On the one hand we show that they are very localized in real space, taking place outside the patches of relative
order, and from the other point of view we show that they represent transitions from one local minimum in the potential surface to another. This picture offers a direct relation between structure and dynamics,
ascribing the slowing down in glass forming systems to the reduction in relative volume of the amorphous material which is liquid-like. While we agree with the well known Adam-Gibbs proposition that the slowing down is due to an entropic squeeze (a dramatic decrease in the number of available configurations), we do not agree with the Adam-Gibbs (or the Volger-Fulcher) formulae that predict an infinite relaxation time at a finite temperature. Rather, we propose that generically there should be no singular crisis at any finite temperature: the relaxation time and the associated  correlation length (average cluster size) increase at most super-exponentially when the temperature  is lowered. 
 \end{abstract}
\maketitle

\section{Introduction}
In this paper we attempt to provide generic answers to some long-standing questions regarding the
spectacular increase in relaxation times in typical structural glass formers as the temperature is
reduced \cite{96EAN,01Don,06Dyre}. In particular we address the relation between structural and dynamical phenomena, the question of existence of a typical length scale that is associated with the slowing down \cite{05BBBCEHLP,06ABIMPS,07HIMPS,07ILLP}, the nature of ageing in such systems \cite{03BCG} and the issue of existence of a ``true" glass transition temperature $T_g>0 $ where the relaxation time becomes infinite \cite{08EP}. We will base most of our comments on a careful analysis of one model system on which we performed extensive simulations, but will provide supporting evidence also for a second model system and an additional experimental system for which theory had been proposed recently. We will argue that these very different systems share some important characteristics, and we will risk a conjecture that these characteristics are generic for structural glass formers. 
Most importantly, these systems share the characteristic of having, in addition to a crystalline ground state, other states whose energy does not differ much from the ground state, at least for relatively small patches. The existence of such states leads first and foremost to the frustration of
crystallization when the temperature is lowered. Glass formation is accompanied by the creation of local patches, or clusters, of different nature, and these patches are locally stable in the precise sense of having  high local shear moduli. To make this point crystal clear (pun not intended) we will introduce and study the notion of a local shear modulus for systems of this kind. The existence of the inhomogeneity (or patchiness) is characterized by a typical scale, and we will demonstrate how this typical scale increases rapidly upon decreasing the temperature. Ageing and relaxation events occur generically in the remainder of the system, in regions of low local shear moduli, as will be shown below. The squeezing out of the regions where relaxations
occur is the fundamental reason for the slowing down, and we connect the increase of typical scale
with the increase of relaxation time. We will analyze individual ageing events and show that they
can be understood either as localized  ``excitation chains" \cite{06Lan} in real space or transitions between local minima of the potential surface \cite{82SW}. Finally we will argue that our picture does not predict the existence of a `true' glass transition, in the sense that infinite relaxation times are expected only as $T\to 0$ (unless a very unlikely catastrophic crystallization event intervenes at a finite temperature).

In Section \ref{binary} we present new results for the binary mixture glass-forming model, studying in a variety of ways, qualitative and quantitative, the spatial inhomogeneity that is crucial for glass formation.
In Section \ref{local}  we introduce the notion of ``local shear modulus", explain how to compute it in numerical simulations, and demonstrate the utility of this notion. We show that in the clusters of relative order  the local shear modulus is typically high, whereas in between the clusters the local shear modulus can be low, even negative, signifying local mechanical instability. Ageing events take place spontaneously in the latter regions. In Sec. \ref{anatomy} we consider the anatomy of the aging events,
and show that they are very localized in real space. In terms of the potential energy landscape the aging
events are mapped to transitions between local minima crossing saddles of order 1, (not necessarily always reducing the potential energy!). In Sect. \ref{models} we discuss the generality of the picture emerging here with the help of other examples of glass formers. Sect \ref{summary} offers a summary and a discussion.
\section{Clusters and heterogeneity in a Binary Glass Former}
\label{binary}

The $2D$ system consists of an equimolar mixture of two types of
point particles interacting via an inverse power potential \cite{89DAY,99PH}:

\beq
\phi_{ab}(r)=\epsilon \left( \frac{\sigma_{ab}}{r} \right)^{n}, 
\hspace{5mm} a,b=1,2.
\label{e1}
\eeq
where $r$ is the separation distance between particles of types $a$ and 
$b$ and $n=12$.
The particles have the same mass $m$, but half of the particles are 'small' 
with 'diameter'  $\sigma_{11}=\sigma$, and half of the particles are 'large' 
with 'diameter' $\sigma_{22}=1.4\sigma$; the interaction between different
kinds of particles is defined by $\sigma_{12}=(\sigma_{11}+\sigma_{22})/2$.
We choose the units of mass, length and time as $m$, $\sigma$ and 
$\tau=\sigma\sqrt{m/\epsilon}$ , respectively. Other reduced units are the system energy $U^{*}=U/\epsilon$,  enthalpy $H^*=H/\epsilon$,  pressure
$P^{*}=P\sigma^{2}/\epsilon$, reduced temperature $T^{*}=k_{B}T/\epsilon$, 
with $k_B$ being Boltzmann's constant, and density 
$\rho^{*}=N/A^*$, where $N$ is the particle number in the simulation box of dimensionless area 
$A^*=A/\sigma^2$.

We performed Molecular Dynamics simulations in the canonical (NVT) ensemble
with $N=1024$ particles in a square simulation box of side $L^*=\sqrt{A^*}$ 
with periodical boundary 
conditions. The equations of motion  were integrated using a third order Gear 
predictor-corrector algorithm \cite{AT} with 
a time step $\delta\tau=0.005\cdot\tau$. A constant temperature was 
preserved using a velocity rescaling method \cite{Heerman}. At each 
temperature the density was chosen in accordance with the simulations results 
in an NPT ensemble as described in \cite{99PH} with the pressure value fixed at 
$P^{*}=13.5$.

\subsection{Clusters and heterogeneity - qualitative picture}
\subsubsection{Voronoi diagrams}
The Voronoi tessellation is an effective tool for the structural analysis of 
many-body systems. In the particular case of two-dimensional systems 
the Voronoi polygon construction associates with every configuration of the
particles a subdivision of position space into polygons, one per particle. The
polygon associated with any particle contains all points closer to that
particle than to any other particle. The edges of such a polygon are the
perpendicular bisectors of the vectors joining the centers of neighboring
particles. 

\begin{figure}
\centering
\includegraphics[width=0.50\textwidth]{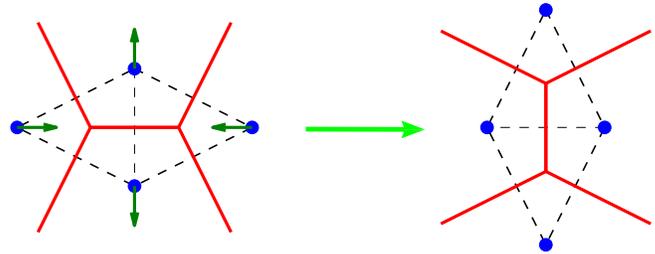}
\caption{ A T1 process in the Voronoi tessellation (in red lines).}
\label{T1}
\end{figure}

Any Voronoi construction obeys Euler's theorem : $\chi=V-E+F$, where
$\chi$ is the Euler characteristic, $V$, $E$ and $F$ are respectively the
number of vertices, edges and faces. In an infinite two-dimensional
system with periodic boundary conditions (on the torus $\chi=0$),
the average number of sides of the polygons is exactly 6 \cite{JSt}.  
This fixed average should be conserved under a local rearrangement of the
polygons. The elementary collective movement of particles which leads to local
topological rearrangement of polygons, and known as $T1$ process, is shown in
Fig. \ref{T1}.
 In a binary mixture there are two types of particles (small and large, or blue 
and red, respectively). Thus to achieve a mapping of the particle positions 
and the number of sides of the polygons in the Voronoi tessellation, we 
distinguish between polygons having a small or large particles in their 
center. Thus, a coloring scheme of cells will take into account not only 
the number of sides, but also the type of the particle in each cell.

\subsubsection{The competing phases}

\begin{figure}[!h]
\centering
\includegraphics[width=0.40\textwidth]{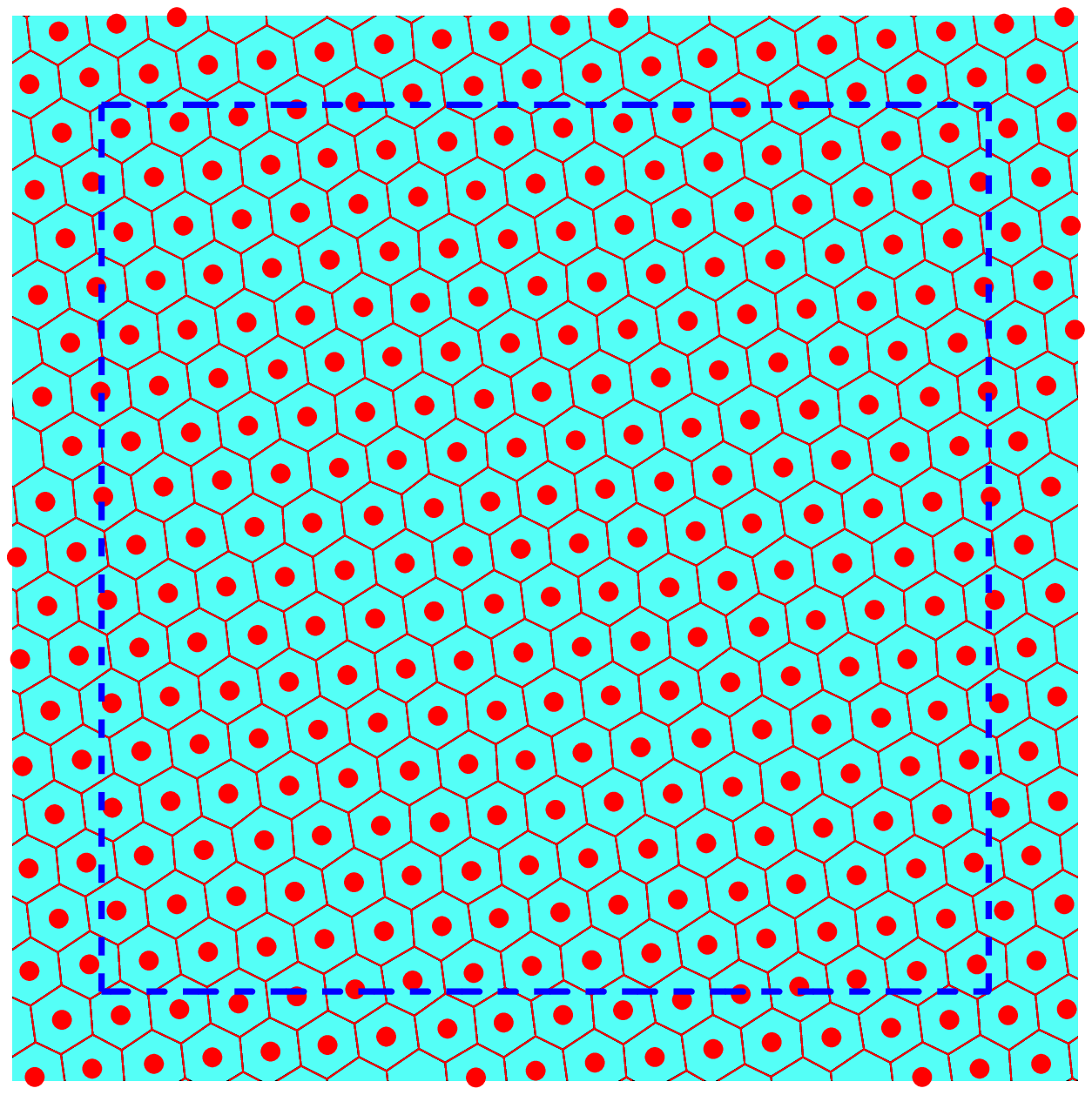}
\includegraphics[width=0.40\textwidth]{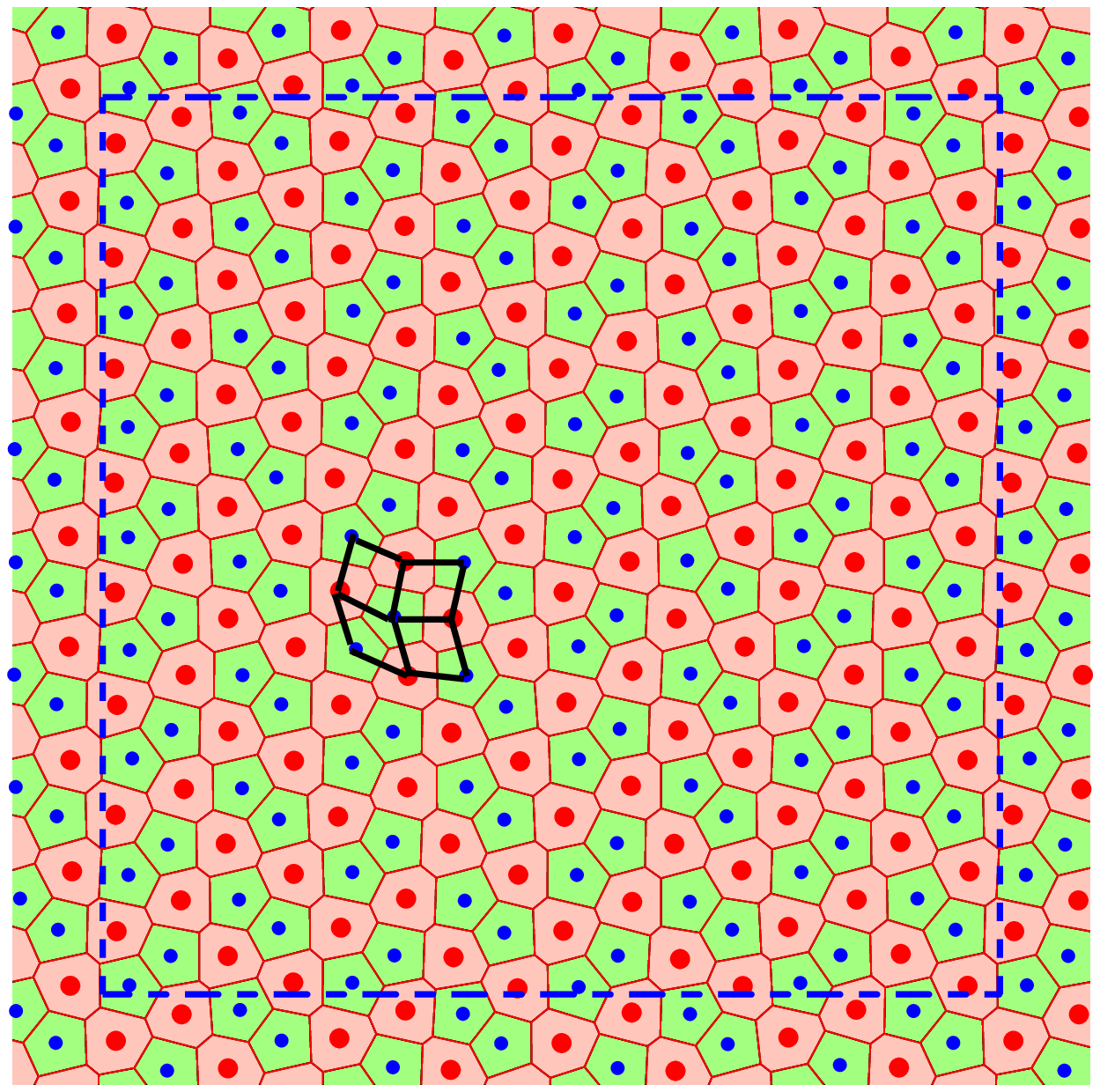}
\caption{Upper panel: a Voronoi polygon construction for the hexagonal 
structure of one-component system. Lower panel: another phase of the binary 
mixture that can be stable at low 
temperatures, made from
a Voronoi tessellation of heptagons with large particles and pentagons with 
small particles.}
\label{qqc}
\end{figure}

At this point we would like to explain that the present system has one ground state,
that of phase separated hexagons of small and large particles, but nearby there
is another homogeneous state which is not too far in energy.
We expect that the ground state of the system under consideration 
should be homogeneous. The obvious candidate structure is a phase-separated 
large and small particles  (an example of Voronoi tessellation for 
subsystem of small or large particles  is shown in the upper panel of 
Fig. \ref{qqc}). But this is not the only possible homogeneous phase; It was shown  \cite{CI95,LSW01,06ABIMPS} 
that the phase shown in the lower  panel of Fig. \ref{qqc} is possible for 
our binary systems in the range $1.18<\sigma_{22}/\sigma<2$. This phase consists of small particles 
in pentagons and large particles in heptagons; these are the polygons which 
 in the theory of  two-dimensional melting of one-component 
hexagonal crystals are referred to as  'defects' \cite{TC82}.
This phase is characterized by  pairs of small particles in 
pentagons lying in parallel lines \cite{CI95}. In one line all the pairs have the 
same orientation but can change it from line to line. Therefore, this phase is 
periodic in one direction and can be ordered or disordered in the other one.
It is interesting to compare these two homogeneous states in terms of their
energy and enthalpy, since we will argue below that they compete in the glassy state.

The ground state of this model (at $T=0$) is defined by minimum of the enthalpy 
$H^*=U^*+P^*\cdot A^*$ or the energy $U^*$ in an NPT or NVT ensemble respectively.
In the appendix we show that the ground state of our system is indeed the phase separated
hexagons of small and large particles. The phase consisting of pentagons and heptagons is
however close in energy, and certainly once formed will not be easy to deform to the 
phase separated ground state.

\begin{figure}[!h]
\centering
\epsfig{width=.4\textwidth,file=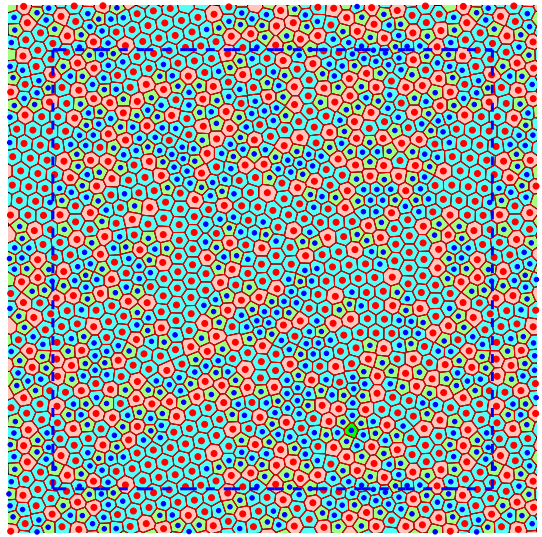}
\caption{A Voronoi tesselation of a typical glassy state. Note the existence of clusters of large particles
in hexagons, small particles in hexagons and patches of the phase shown in Fig. \ref{qqc}. Our point of view is that glass formers generically form  clusters of such competing ground and close to ground states.}
\label{clusters}
\end{figure}

The crucial comment is that in the super-cooled state this model exhibits clusters of all these phases.
In  Fig. \ref{clusters} we show snapshots of the system at the temperature 
$T^*=0.2$.   In the remainder of this paper we connect between the heterogeneity that is caused by
the existence of such clusters and the phenomenon of slow ageing in the glassy state.    
Ref. \cite{99PH} found, using molecular dynamics simulations in the 
isothermal-isobaric ($NPT$) ensemble, that for temperature $T>0.5$ the system 
is liquid and  for lower temperatures dynamical relaxation slows down. A 
precise glass transition had not been identified in \cite{99PH}. In 
\cite{06ABIMPS,07HIMPS} it
was argued on the basis of statistical mechanics that there exists a typical 
length scale that
grows super-exponentially fast when the temperature decreases. Associated with 
this fastly growing scale there exists a fastly growing relaxation time, such 
that below a certain temperature the system is jammed for all practical 
purposes. Here we will shed further light on this phenomenon, relating it to 
the inhomogeneity seen with the bare eye in Fig. \ref{clusters}. 

\section{Clusters and heterogeneity- quantitative picture using the local shear modulus}
\label{local}

In an NVT ensemble a given configuration can be characterized by components
of the microscopic stress tensor which is defined by (see, e.g. \cite{Crox74}):
\beq
\sigma_{\alpha\beta}=\frac{1}{A^*}\left(\sum\limits_{i} 
p^{\alpha}_ip^{\beta}_i-\frac{1}{2}\sum\limits_{j\ne i}
\frac{\partial \phi_{ab}(r_{ij})}{\partial r_{ij}}
\frac{r^{\alpha}_{ij} r^{\beta}_{ij}}{r_{ij}}\right),
\label{str}
\eeq
where $p^{\alpha}_{i}$ is the $\alpha$ component of the dimensionless momentum
of particle $i$ and $r^{\alpha}_{ij}$ is the $\alpha$ component of the vector 
joining particles $i$ and $j$.

The diagonal components of this tensor correspond to density fluctuations and its trace
determines the pressure, in accordance with the virial theorem:
\beq
P^*=\rho^* T^*-\rho^*\frac{1}{2d}\frac{1}{N}\sum\limits_{j\ne i}
r_{ij}\frac{\partial \phi_{ab}(r_{ij})}{\partial r_{ij}},
\label{virial}
\eeq
where $d$ is the spatial dimension.
[h]
\begin{figure}
\centering
\epsfig{width=.30\textwidth,file=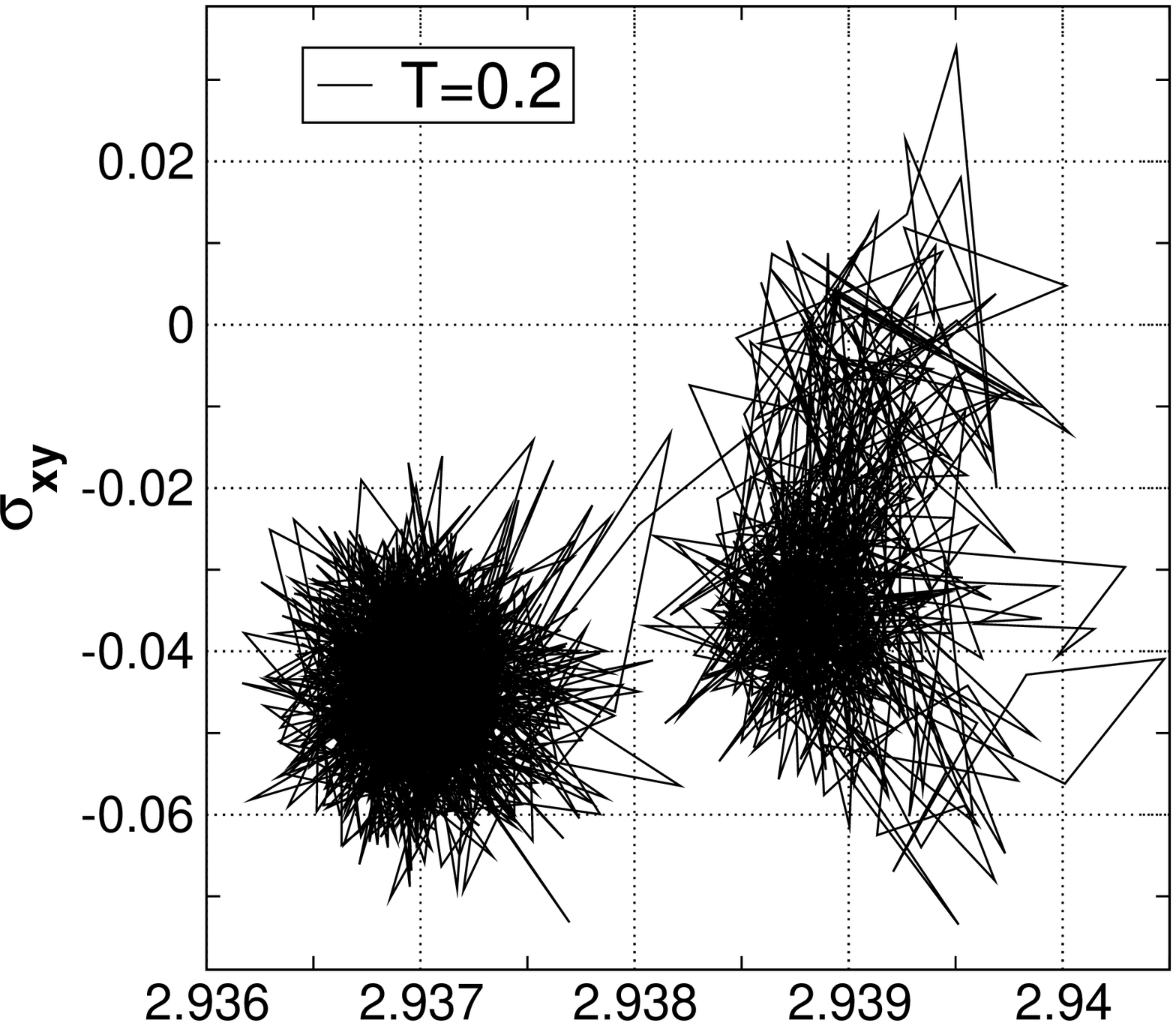}
~\vskip 1 cm
\epsfig{width=.33\textwidth,file=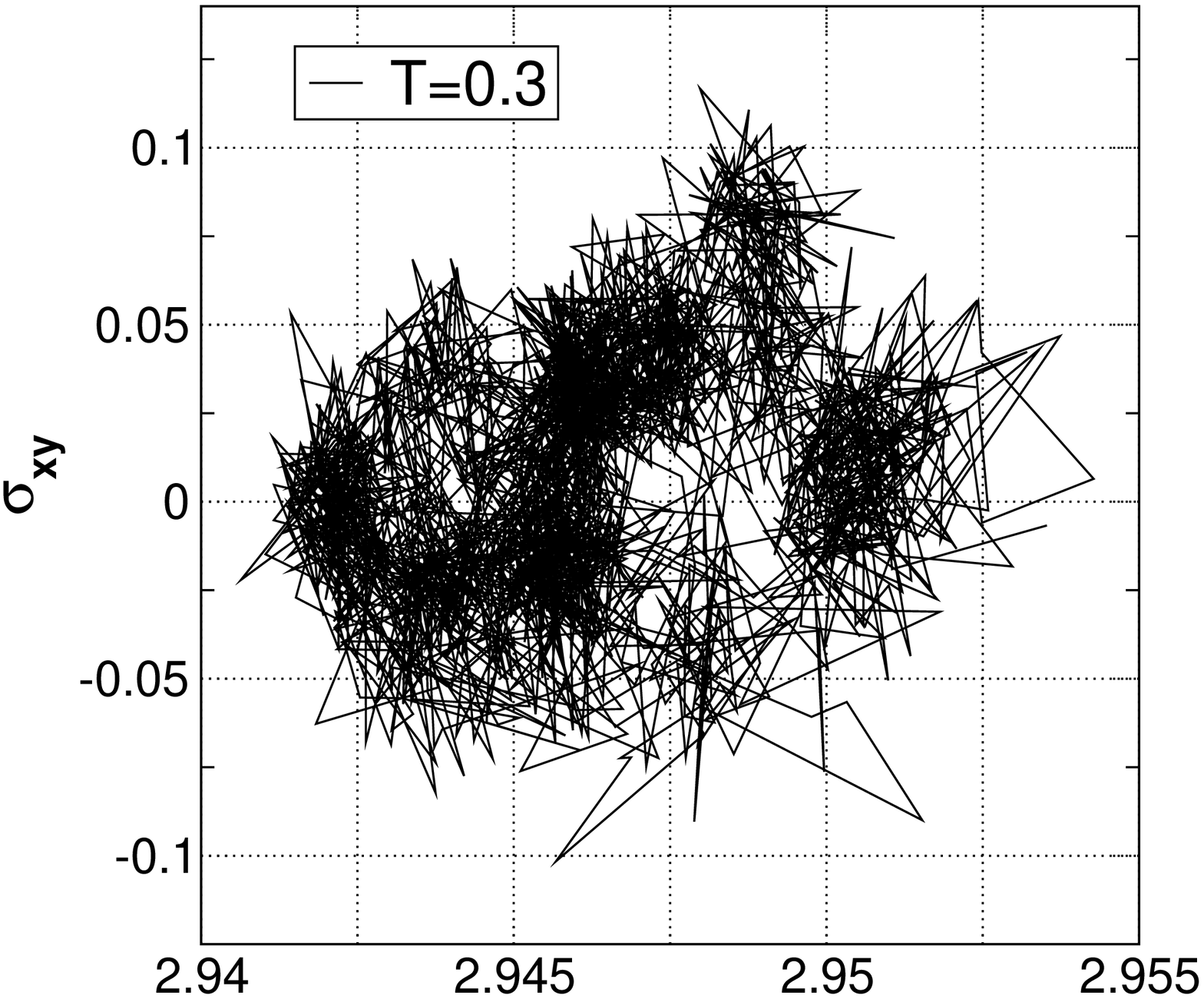}
~\vskip  1 cm
\epsfig{width=.30\textwidth,file=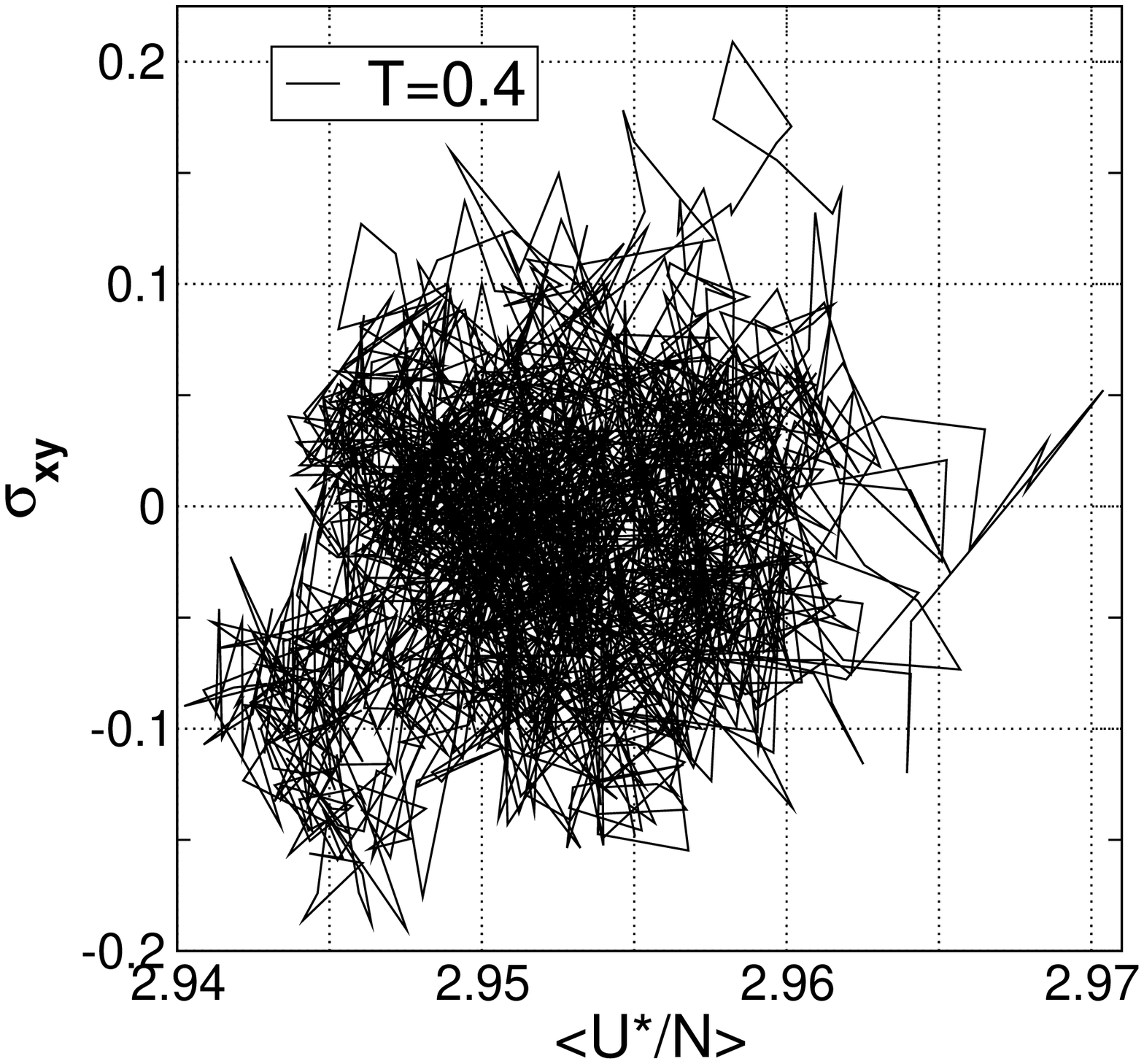}
\caption{Trajectories from Molecular Dynamics simulations in
  the stress-energy plane, for the three temperatures T=0.2, T=0.3 and T=0.4 respectively. $ U^*/N$ is
  the potential energy of the system per particle. Each point represents an 
average over a duration of 5$\tau$, connected with straight lines. The time 
of simulation  is $10^4\tau$. Note the dramatic increase in the scale of the variations when the temperature increases.}
\label{jumps}
\end{figure}
For interactions of the form (\ref{e1}) the pressure is related
to the total potential energy of the system (\ref{engen}):
\beq
P^*=\rho^* T^*+\rho^*\frac{n}{d}\frac{U^*}{N}.
\label{pu}
\eeq
The point is that an economic designation of a `state' is given  in
the NVT ensemble by the pair of quantities 
energy and shear stress $\sigma_{xy}$.

In Fig. \ref{jumps} we exhibit the actual trajectory of the Molecular dynamics 
simulations in the stress-energy plane for three
different temperatures and for the same simulation time. We see that at 
$T=0.2$  the trajectory hovers mostly around two distinct states with infrequent
transitions between them (only one transition in this run). This is a general observation
for low temperatures: the system fluctuates 
around one  ``solid-like'' state, but then jumps to another such 
``solid-like'' state. The ease of transitions increases with temperature but also with the number of
particles. Here we have 1024 particles; in  previous simulations using 256 particles \cite{07IMPS} we did not observe any transition at $T=0.2$.  Changing the temperature to $T=0.3$ but keeping the same simulation time one resolves  seven-eight ``solid-like'' states.  For $T=0.4$ the trajectory now 
fills up an extended region in the stress-energy plane. As stated in \cite{07IMPS}, it appears
that between `real' solid and `real' 
liquid the system is locally not a ``viscous fluid''. Rather, the trajectory 
concatenates relatively long period of time where the system behaves like a 
solid, interconnected by relatively short period of times where
the system flows between these states. Clearly, a viscous fluid behaves very 
differently, responding
to stress by a viscous flow, be it as sluggish as one wishes. Here, most of 
the time, the system does not
respond to stress, except in the narrow corridors of flow which become rare 
when the temperature goes down and more common when the temperature warms up. 
Of course this does not mean that in the sense of long time averaging the 
notion of viscosity cannot be resurrected, but locally in time this
is impossible.

\begin{figure}
\centering
\includegraphics[width=0.55\textwidth]{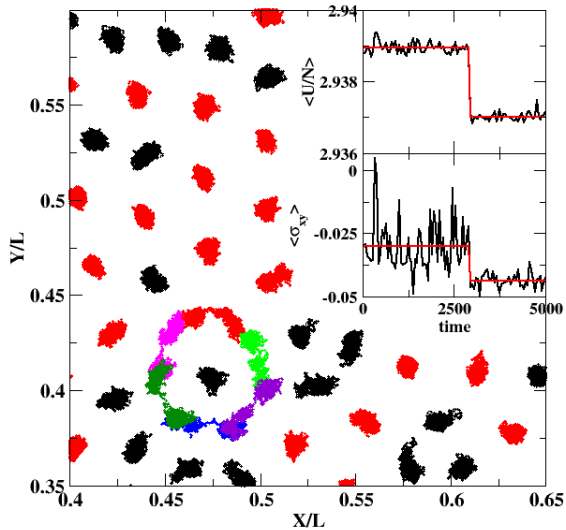}
\caption{Details of an event of the type seen in Fig. \ref{jumps} leading to 
a jump from one
solid-like state to another. In the inset we show the energy per particle and 
the average shear stress  as a function of time. Every point in these graphs 
is an average over $10^4$ simulation steps. The figure itself
is a window in the simulation box showing that the event is a circular motion 
of particles replacing
each others position. The color code outside the event is red particles 
are large and black are small. In the circular event the red particle is large 
and all the rest are small }
\label{event}
\end{figure}

Since we have the system trajectory at our disposal, we can examine more 
closely the structural
transformation that is taking place when the system jumps from one 
``solid-like'' state to the other. In Fig. \ref{event} we show what happens by 
monitoring three characteristics of the system.
One is simply the energy per particle as a function of time cf. upper inset. 
We see that the energy fluctuates around a given value, and then jumps to a 
new state where the energy fluctuates at
a lower value. A second characteristic is the average shear stress 
$\langle \sigma_{xy}\rangle$ 
(cf. lower inset) which exhibits a similar jump, except that the variance of 
its fluctuations before
the transition appears considerably larger than the same variance after the 
jump. Most interestingly, the figure shows a window into the particle 
configuration itself, focusing on the local event that is
responsible for the jumps in the insets. We see that the event is a concerted 
motion of six particles that change positions along a circular path; such events are referred
to as ``excitation chains" \cite{06Lan}. At this point we want to explore 
what is the significance
of this event, why it takes place where it does, and what is it accomplishing 
in terms of ageing and
relaxation. To this aim we will study local mechanical properties.

\subsection{Local shear moduli}

For inhomogeneous systems such as amorphous materials it is important to study
local properties in connection with thermodynamical and structural
characteristics. To measure local values we take our simulation box and
subdivide it into smaller squares of length $l_{k}=L^*/2^k $ ($2\le k \le 4$).
We calculated the local stress and the local shear modulus 
following the definition of these quantities in \cite{lutsko88}. In the frame
of this approach the local stress tensor ascribed to a subdomain $m$ ($1\le
m\le k^2$ in the $k$th level) is given by \cite{{lutsko88},{YJWNP04}}:
\beq
 \sigma^{m}_{\alpha\beta}=\rho^*_{m}T^*\delta_{\alpha\beta}-\frac{1}{2\cdot
 l_{k}^2} \sum\limits_{j\ne i}
\frac{\partial \phi_{ab}(r_{ij})}{\partial r_{ij}}
\frac{r^{\alpha}_{ij} r^{\beta}_{ij}}{r_{ij}} \frac{q^{m}_{ij}}{r_{ij}},
\label{lstr}
\eeq  
where $\rho^{*}_{m}$ is the particle number density in the subdomain $m$, 
and $q^{m}_{ij}$ is the length of the line segment of the vector joining 
particles $i$ and $j$ which is located inside a subdomain $m$ (if this vector 
does not pass through this subdomain, $q^{m}_{ij}=0$).
The term $q^{m}_{ij}/r_{ij}$ determines how different pairwise
interactions are apportioned to the local stress of a subdomain $m$, and
includes contributions from pairs of segments located outside the
subdomain. An average of $\sigma^{m}_{\alpha\beta}$ over the entire area of
the simulated system yields the usual bulk stress tensor (\ref{str}).

\begin{figure}
\centering
\includegraphics[width=0.40\textwidth]{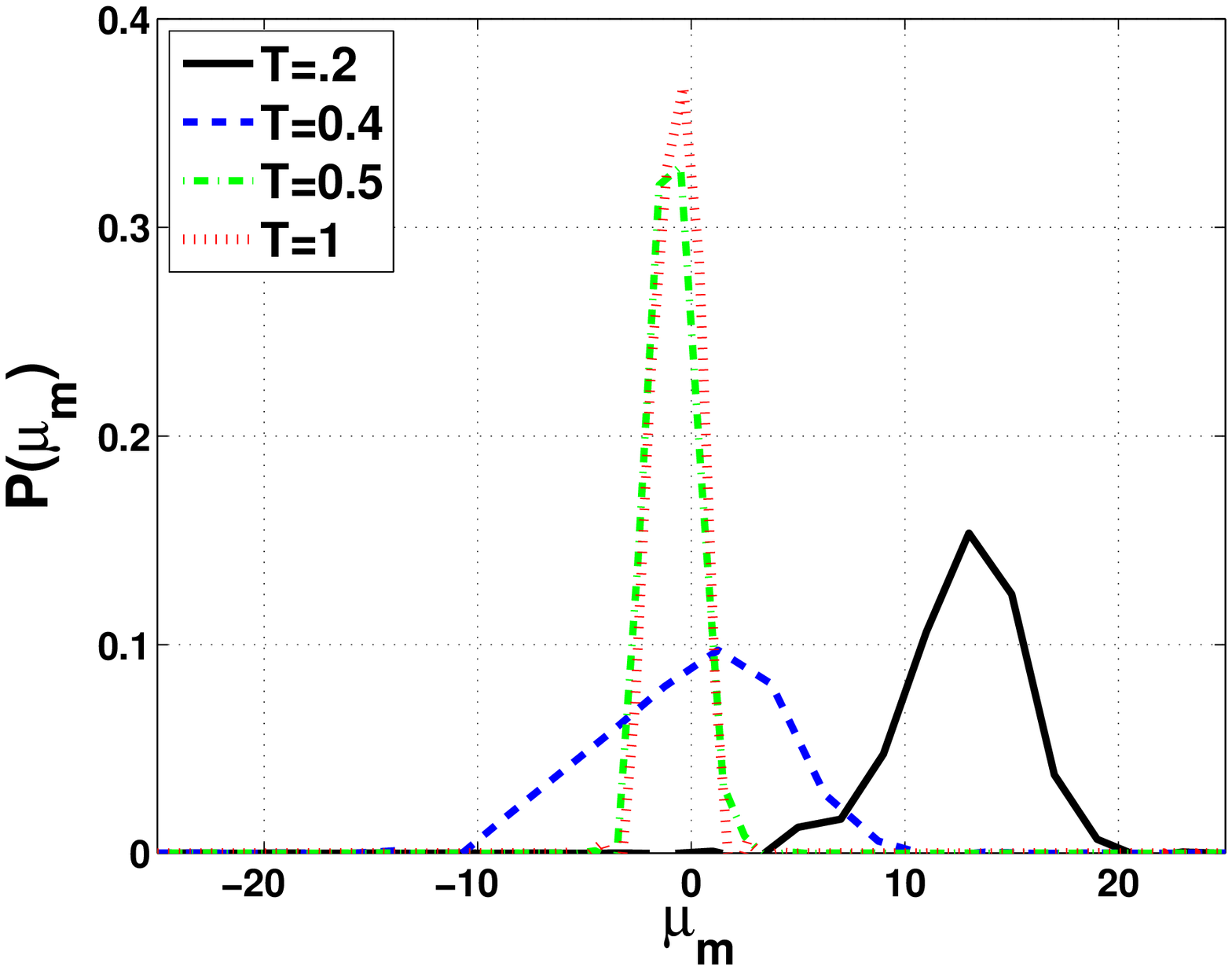}
\includegraphics[width=0.40\textwidth]{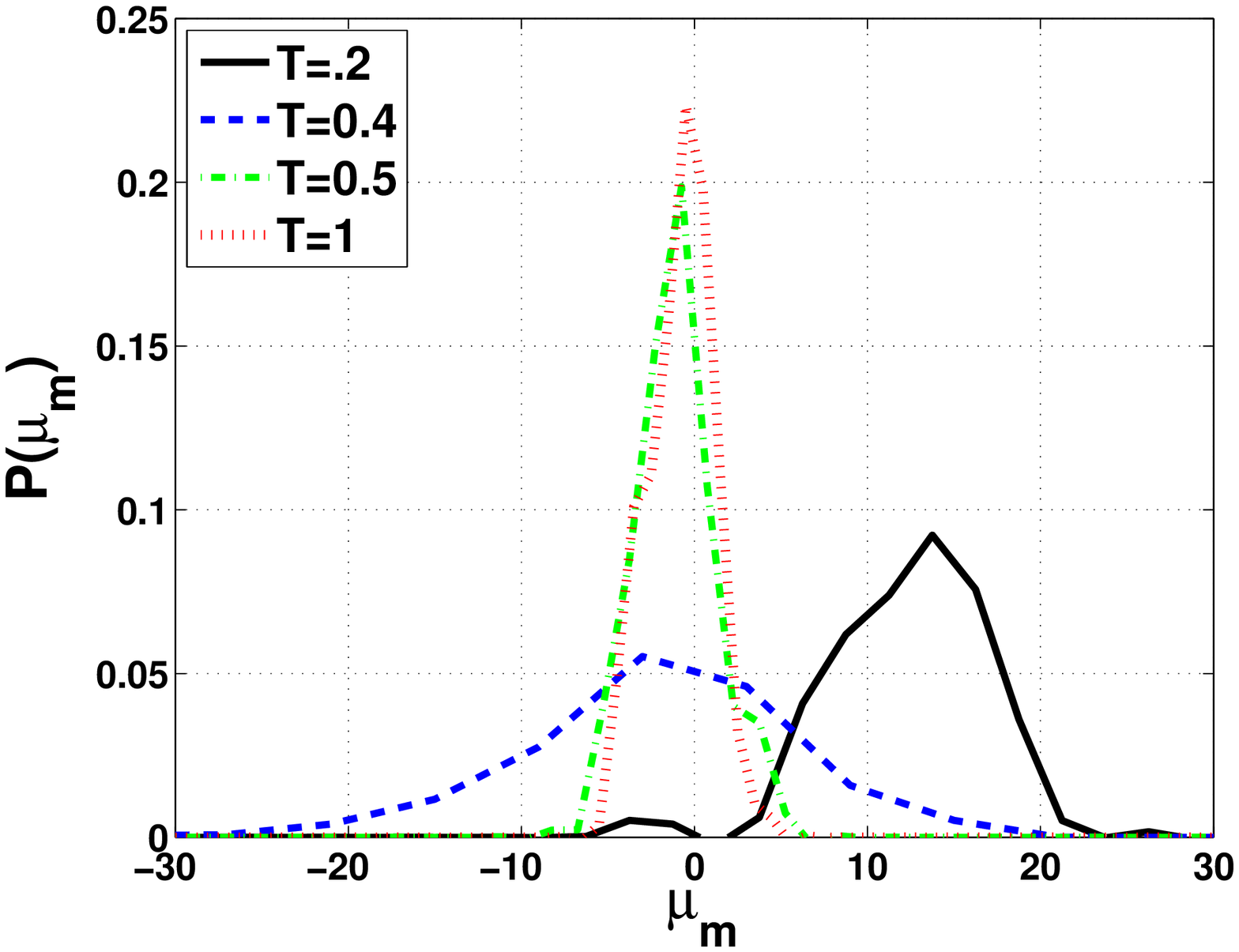}
\includegraphics[width=0.40\textwidth]{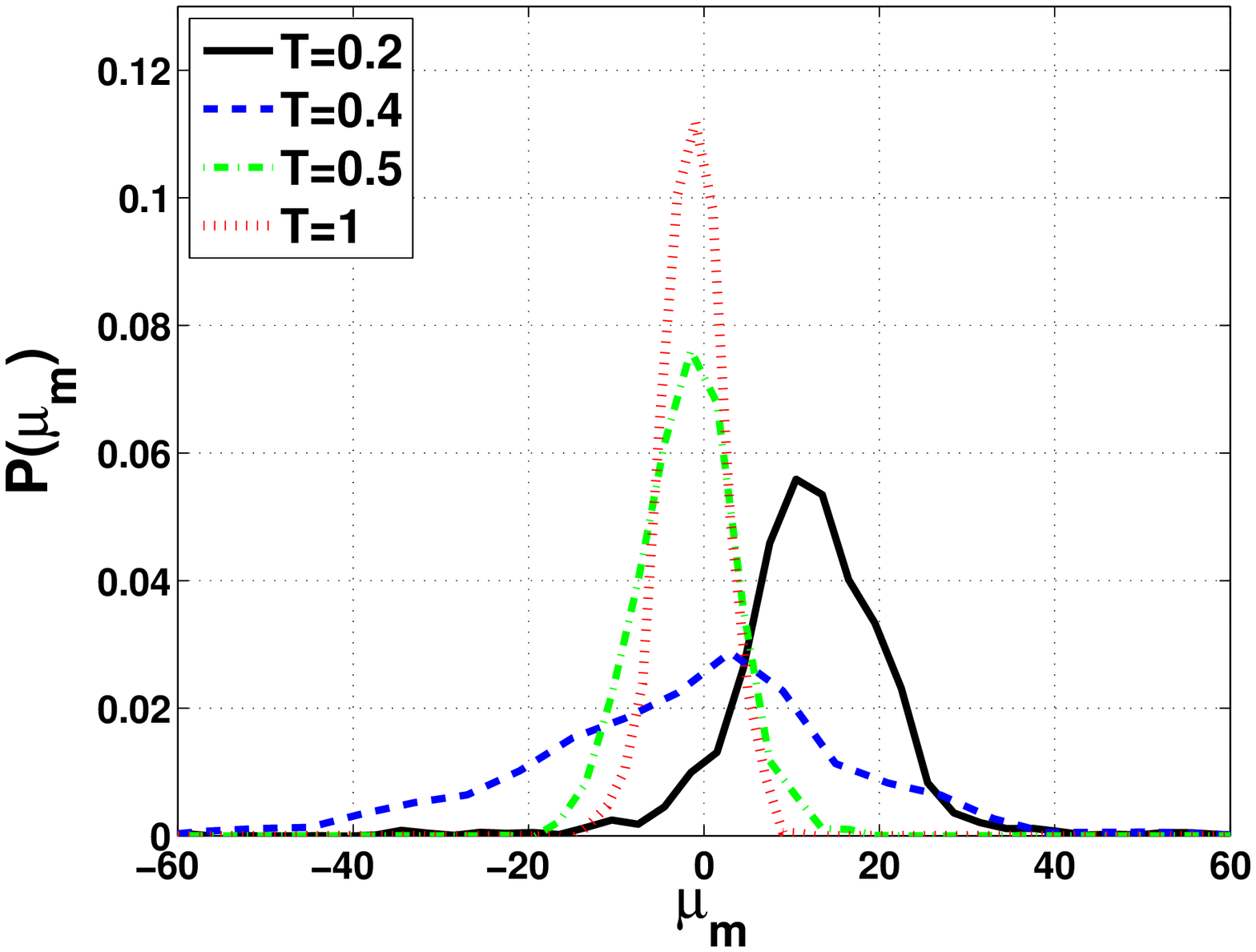}
\caption{ Histograms of the local shear moduli for different temperatures (see the
line-type in the inset) and for different partitions, from gross ($k=2$, upper panel) to fine
($k=4$, lower panel). Note the tendency to move the average to positive values with lowering the temperature, with nevertheless a tail at $T=0.2$ going into negative values of the local shear modulus.}
\label{hde}
\end{figure}

The local elastic modulus tensor is related to the internal stress
fluctuations. The expression for the shear modulus per subdomain is defined by 
\cite{{lutsko88},{YJWNP04}}:
\beq
\mu_{m}=\mu^{\infty}_{m}-\frac{{L^*}^2}{T^*} \left( \langle\sigma^{m}_{\alpha\beta} 
\sigma_{\alpha\beta}\rangle-\langle\sigma^{m}_{\alpha\beta}\rangle \langle\sigma_{\alpha\beta}\rangle\right),
\label{mod}
\eeq
where angular brackets mean averaging over configurations and the ``infinite 
frequency shear modulus" $\mu_m^\infty$ \cite{65ZM,00Wal} per subdomain is given
by:
\begin{eqnarray}
\mu^{\infty}_{m}&=&\rho^*_{m}T^* \label{modinf}\\
&+&\frac{1}{2 l_{k}^2}
\Bigg\langle\sum_{i\ne j}\Bigg[ r\frac{\partial}{\partial r} 
\Bigg[\left(\frac{1}{r}\frac{\partial \phi_{ab}(r)}{\partial r}\right)\!
\!\left(\frac{r^{\alpha} r^{\beta}}{r^2}\right)^2 
\!\frac{q^{m}}{r}\Bigg]_{ij}\Bigg\rangle \ . \nonumber
\end{eqnarray}
\subsection{Results}

The results for the local shear modulus can be exhibited in a number of forms, each shedding
light on another aspect of this useful measure. In Fig. \ref{hde} we show histograms of the
local shear modulus, for different levels of box resolution, from coarse to fine as one goes
from the upper panel to the lowest.  In general we see that as long as the temperature is high
such that the system is fluid, the histograms peak around zero and are quite narrow. With $T=0.4$ the histogram widens, but the mean still moves only a bit from zero. Not so for $T=0.2$ which is
deep in the glassy domain, where the histograms move to exhibit a positive average, but still
showing wide variations in the values of the local shear modulus. Note that with the increase of
resolution we find more negative values of the shear modulus, and these are indicative of
sensitive regions in the system which are mechanically ready to exhibit ageing events. To 
clarify this point further  we turn now to another way of presenting the same results, color coding real space to exhibit the high degree of spatial inhomogeneity of this system. 

\begin{figure}
\centering
~\hskip -0.98 cm
\includegraphics[width=0.60\textwidth]{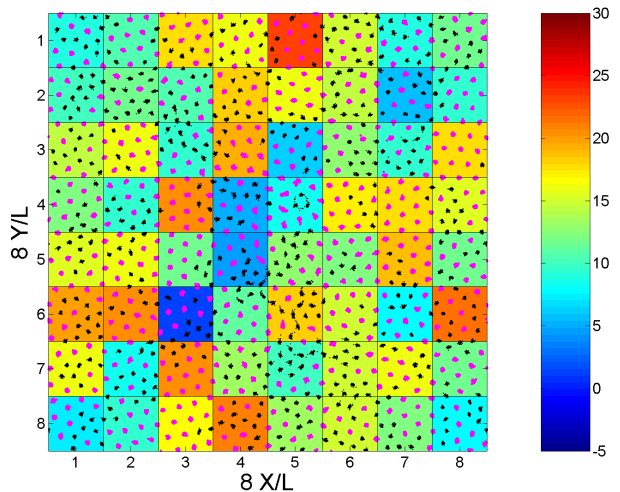}
\caption{Color-coded local shear modulus before the ageing transition. Note that the local values can be either positive or negative, with positive values indicating local mechanical stability, whereas local negative values indicate mechanical instability. We expect relaxation (ageing) events to occur in the blue regions.}
\label{before}
\end{figure}

\begin{figure}
\centering
~\hskip -1 cm
\includegraphics[width=0.60\textwidth]{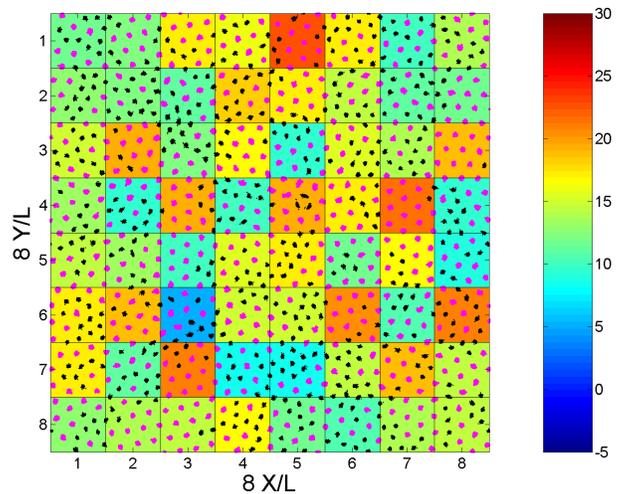}
\caption{Color-coded local shear modulus after the local relaxation event occurred in box (4,5). }
\label{after}
\end{figure}

\begin{figure}
\centering
~\hskip -2.4 cm
\includegraphics[width=0.70\textwidth]{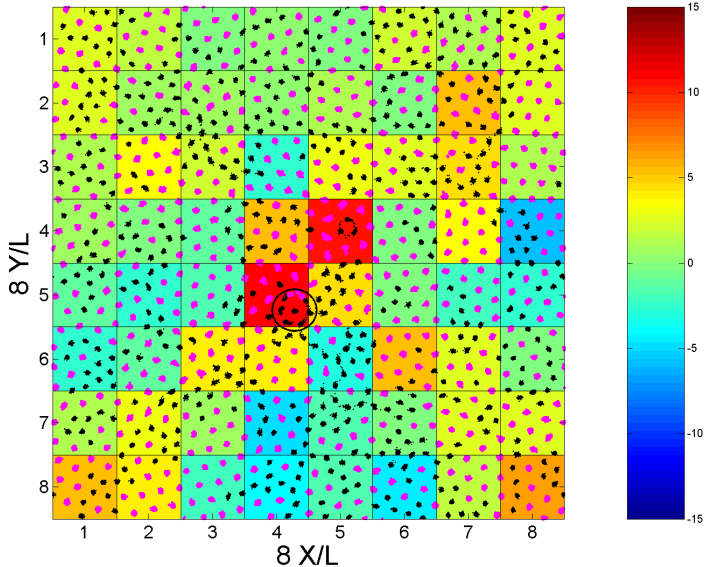}
\caption{Color-coded difference in the local shear modulus between the two last figures, computed after the local relaxation event occurred in box (4,5). }
\label{diff}
\end{figure}

\begin{figure}
\centering
~\hskip -2.4 cm
\includegraphics[width=0.70\textwidth]{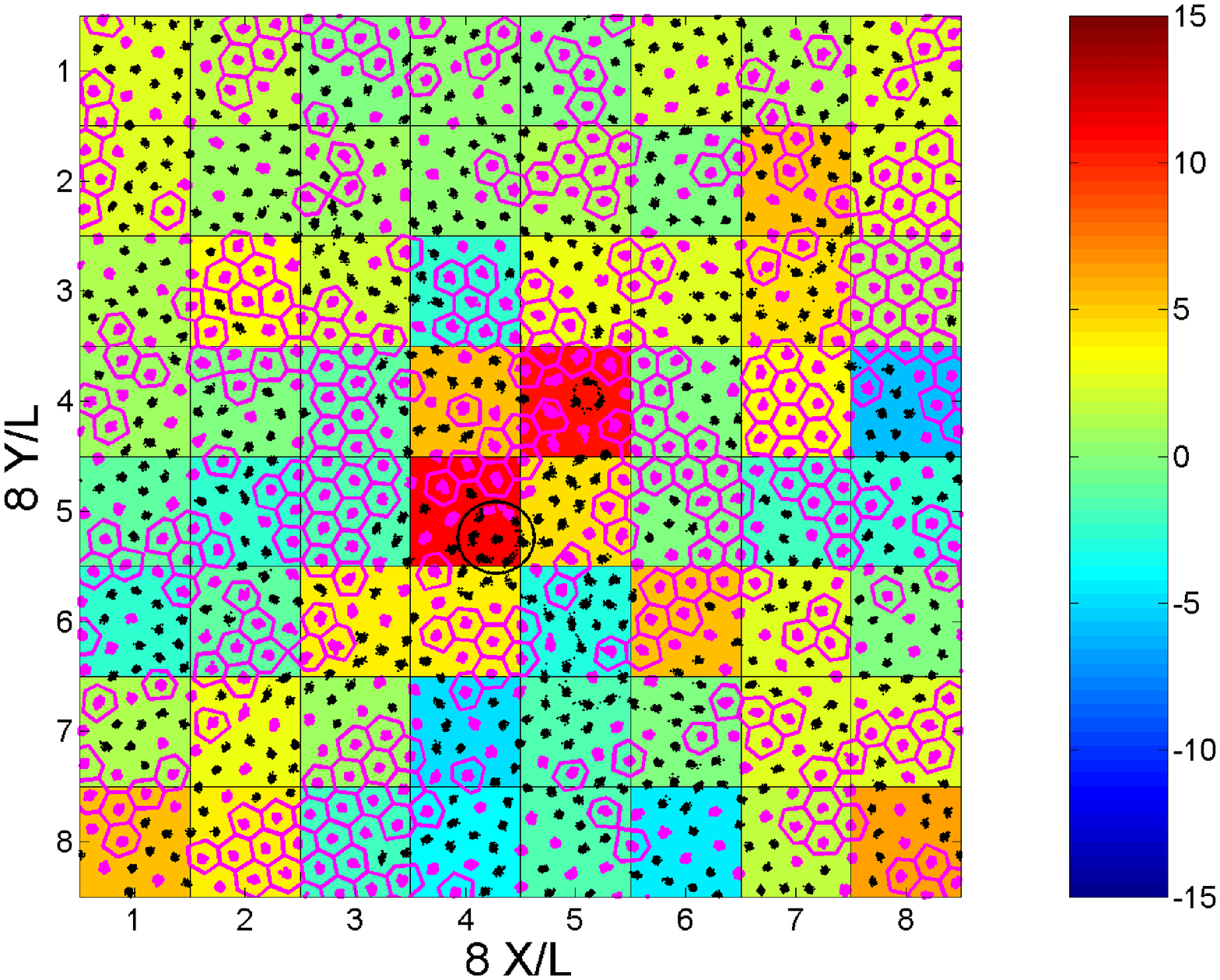}
\caption{The same as Fig. \ref{diff} but with the clusters of large particles superposed to show
that the main changes occur outside the clusters which are locally stable in the mechanical sense.}
\label{with}
\end{figure}

Typical spatial results for the local shear modulus with spatial resolution $k=3$ are shown in Fig. \ref{before}. The local shear modulus is color coded, showing positive and negative values of the local shear modulus. Obviously, in the liquid (high temperatures) the distribution of local values averages to zero when summed over all the cells, whereas when the system
is at low temperatures the average is positive, in agreement with the results of Ref. \cite{07IMPS}. Also, for higher temperatures the spread of values of the shear modulus (or the variance of its distribution) is smaller, whearas the variance also increases upon decreasing the temperature.
We can then expect that spontaneous ageing events will take place where the local shear modulus is low. Indeed, for the system shown in Fig. \ref{before} the first ageing event took place in the square (4,5), which is a region where the local shear modulus attains its lowest value. We checked over many simulations that this is indeed typical, and that looking at the distribution of local shear moduli one can guess very reliably where the next ageing event will take place. Of course, once the ageing event occurred, the distribution of local
shear moduli changes dramatically. In Fig. \ref{after} we show the distribution of local shear moduli {\em after} the event in square (4,5) took place. We see that the local shear moduli went up considerably in the region of the
event; This is shown quantitatively in Fig. \ref{diff} where the difference between the local shear modulus in Figs. \ref{after} and \ref{before} is presented. We certainly need to understand what is precisely happening here.
A clue to that aim is obtained when we superpose on Fig. \ref{before} the clusters of hexagonal patches of big particles, see Fig. \ref{with}. We see that most of the changes in local shear modulus occur outside the clusters which are locally stable in the mechanical sense.

\section{The Anatomy of ageing Events}
\label{anatomy}
In this section we examine the same localized events of ageing, or further relaxation steps towards equilibrium,
but change the point of view from real-space to the potential energy landscape. We start with introducing
the basic notions.

\subsection{Potential Energy Landscapes}
\begin{figure}
\hskip -.05 cm \epsfig{width=.55\textwidth,file=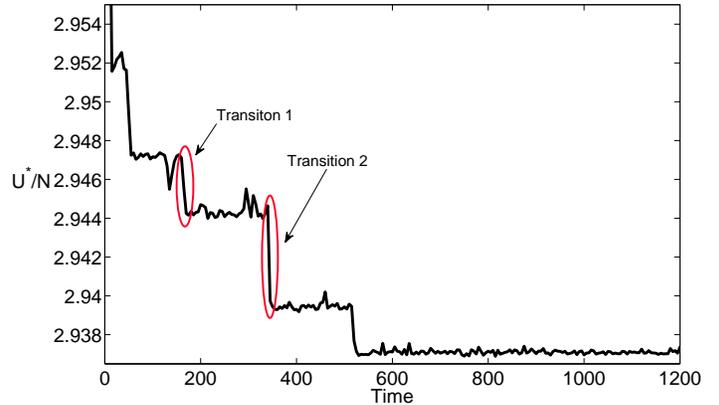} 
\caption{ageing process. Shown is the potential energy per particle measured in the direct
numerical simulation. Note the existence of five relaxation events. We analyze carefully the two events marked with ellipses.}\label{energy}
\end{figure}

One popular way of discussing  the dynamics of many particle systems is to visualize the state of the system as a point on the surface 
that the potential energy function $U = f(\B x)$ draws in a $Nd+1$-dimensional space, where
$N$ is the number of particles and $d$ the dimension of  space ($d=2$ in our examples) \cite{69Gol}. The evolution in time is described by the path of the coordinate vector $\B x$
along the energy surface. Adding the thermal energy to the potential energy elevates the point above the energy surface, but we will be interested in locating the paths associated with the relaxation events on the potential energy surface itself.  At low temperatures the system is trapped in a minimum of this surface. In a crystallized system this minimum corresponds to the global minimum of the system - the ground state. In an amorphous system, however, this minimum is typically not a global one, but rather a local minimum which is separated from other minima of the system by moderate potential barriers. At temperatures greater than zero, we expect that, due to thermal fluctuations, the system will pass from one minimum to another via saddles. These saddles can be
classified by  their `order', which by definition is the number of negative eigenvalues of the Hessian matrix of $U$ at the saddle. At low temperatures one encounters typically saddles of order one \cite{02GCGP} . Our aim here is to
focus on spontaneous ageing events and map them onto trajectories on the energy surface. To do so we will take data from our numerical simulations, locate the ageing events, 
and for each point in time before and after the event find the nearest extremum of the potential. The methods used to find the extrema are standard, and are summarized for convenience in Appendix \ref{extrema}. Unfortunately the numerics involved in this procedure is rather heavy, and for that
reason in this section we will refer to simulations with 256 rather than 1024 particles and also somewhat
higher temperatures where there is a larger number of ageing events, cf Fig. \ref{jumps}. The insights achieved are of course independent of this numerical restriction.
\begin{figure}
\centering
~\hskip -0.3 cm
\epsfig{width=.55\textwidth,file=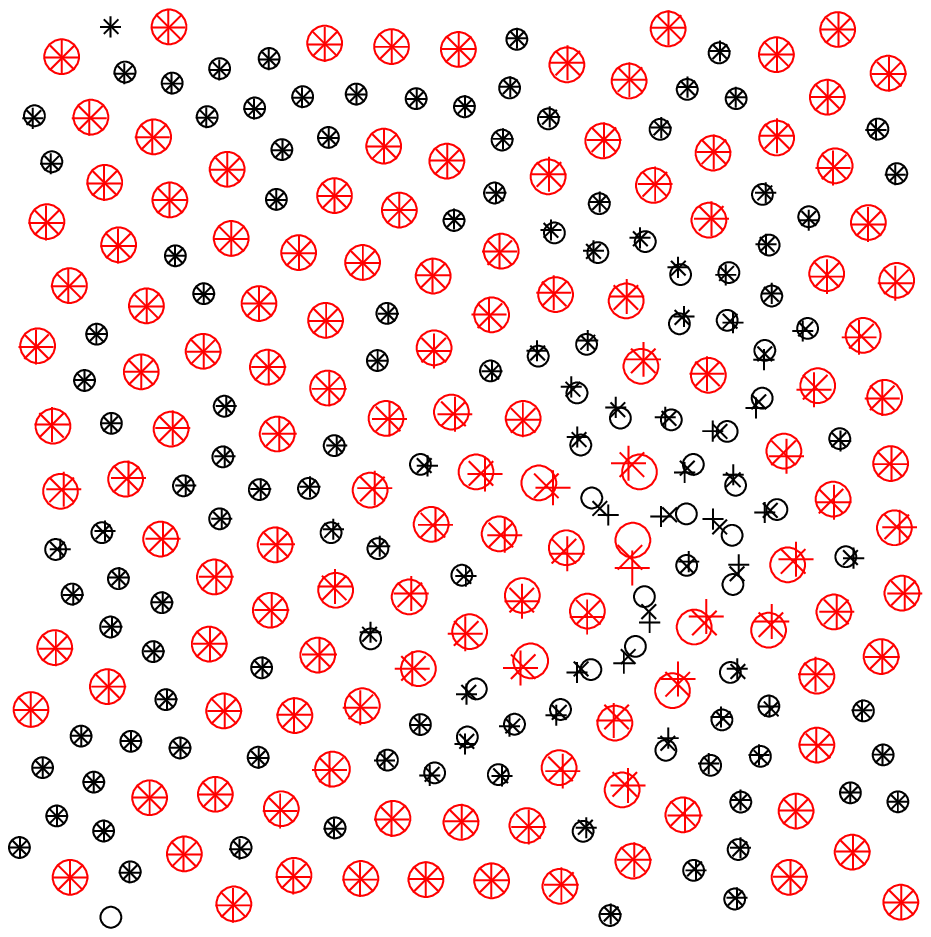}
~\hskip -1.6 cm
\epsfig{width=.55\textwidth,file=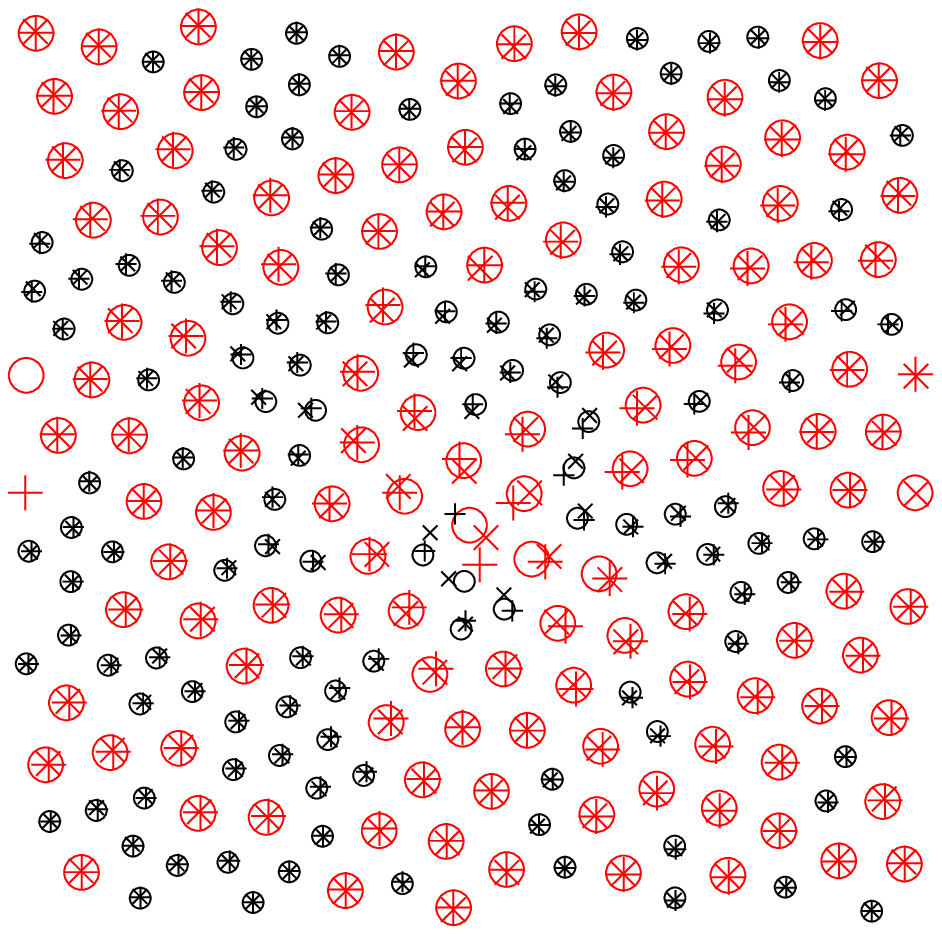} 
\caption{(Color online). Configurational changes during transitions 1 and  2  in Fig.  \ref{energy} respectively. Large symbols refer to large particles and small symbols to small particles. In circles we denote the positions of the particles before the transition, in x the positions at the saddle and in + the positions after the transition. }\label{change}
\end{figure}
\begin{figure}[h]
\centering \epsfig{width=.5\textwidth,file=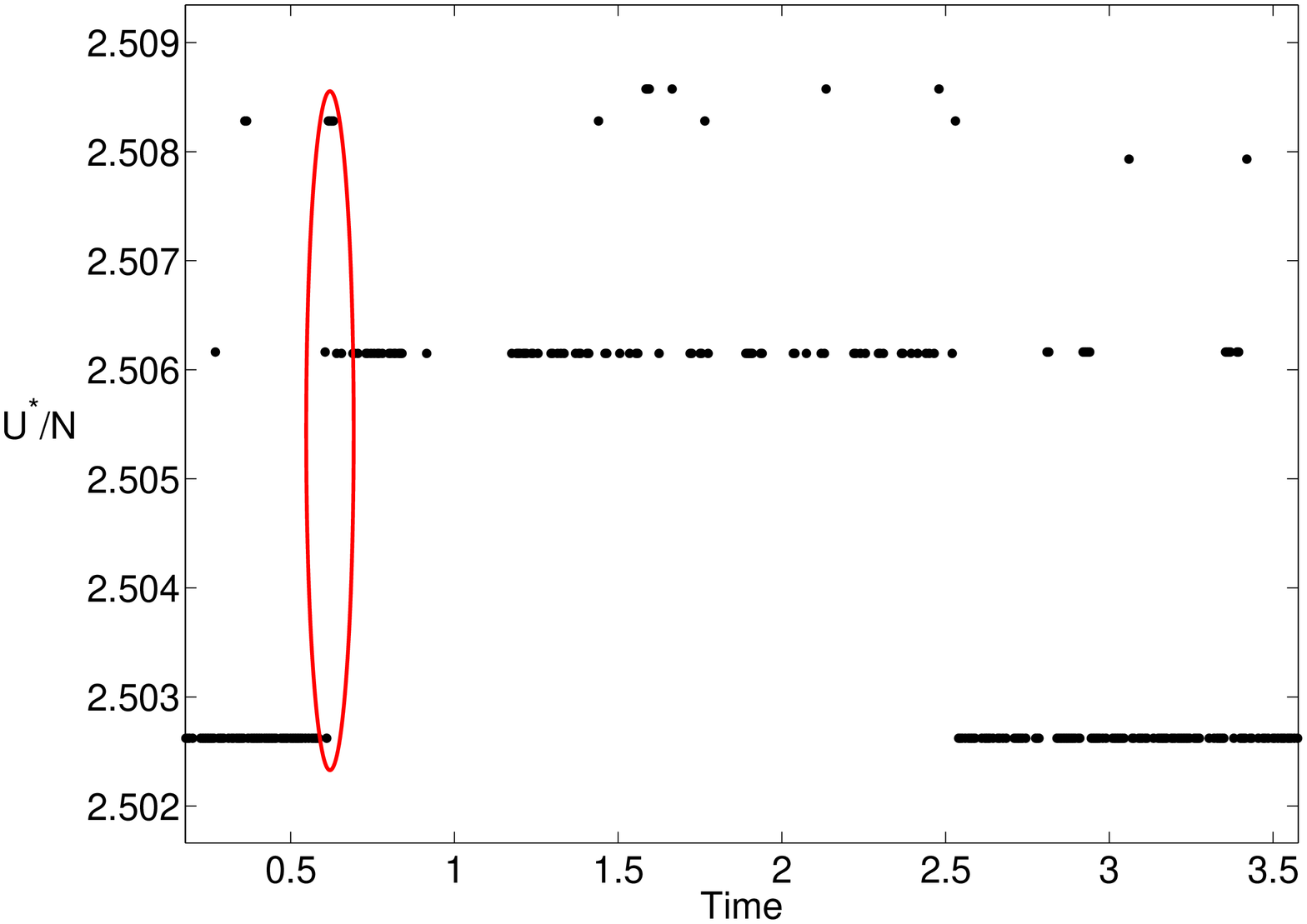} 
 \epsfig{width=.50\textwidth,file=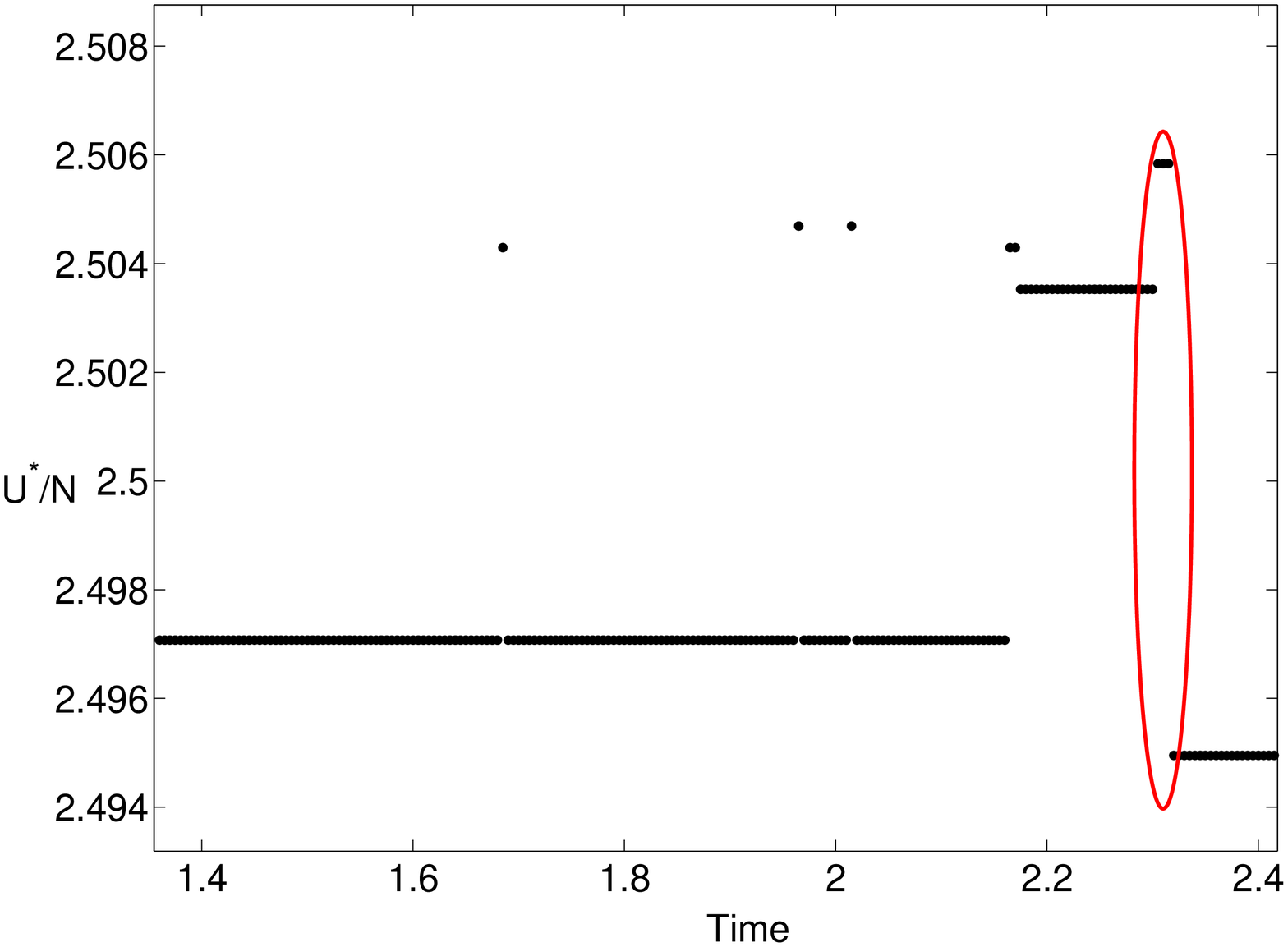} 
 \caption{Energy of nearest extrema as a function of time for the first and the second transitions in Fig. \ref{energy} respectively. Not that in the upper panel we focus on an event in which the energy is increasing, and compare with Fig.~\ref{trans}. }\label{trans6}
\end{figure}
As an example consider 
Fig. \ref{energy} depicting the potential energy per particle as a function of time during an ageing process in a system of 256 particles. This figure contains evidence for five ageing events, and below we analyze in detail  transitions occurring in the temporal proximity of the two events marked in red. These transitions are localized, in the sense that the irreversible change in the
configuration of the system involves only a few particles. 
In Figs. \ref{change},  we can see the actual changes in configuration caused during
these events. Here the particles positions are plotted before, during and after the event. The changed area consists of a mix of small and large particles.  A temporary displacement occurs in a larger area around this region. This collective string-like process \cite{98DDKPPG} allows for the permanent change to take place despite the high pressure. 

Fig. \ref{trans6} shows the energy of the extrema found in the vicinity of these transitions.  To find those we use a time-dependent trajectory from the molecular dynamics simulation and for each
configuration analyzed we seek the closest extremum of the potential energy surface. The vertical scale shows the potential energy of the system at the nearest extremum found using the LBFGS algorithm (see Appendix \ref{extrema}). Most of the time the method locates a particular extremum with a specific energy, starting with a minimum, going through
a saddle and ending at another minimum. We make sure that these extrema are indeed minima or saddles by finding the
smallest eigenvalue of the Hessian matrix at each extremum; it is positive for the minima and unique negative in the saddle, meaning that the saddle is of order unity. 

Note that during the long residence near the first minimum in Fig. \ref{trans6} there are instances when the system fluctuates to a point near a saddle;  However the system remains trapped in the first minimum until the ageing transition takes place. 

\begin{figure}[h]
\centering 
\includegraphics[width=8cm,height= 10 cm]{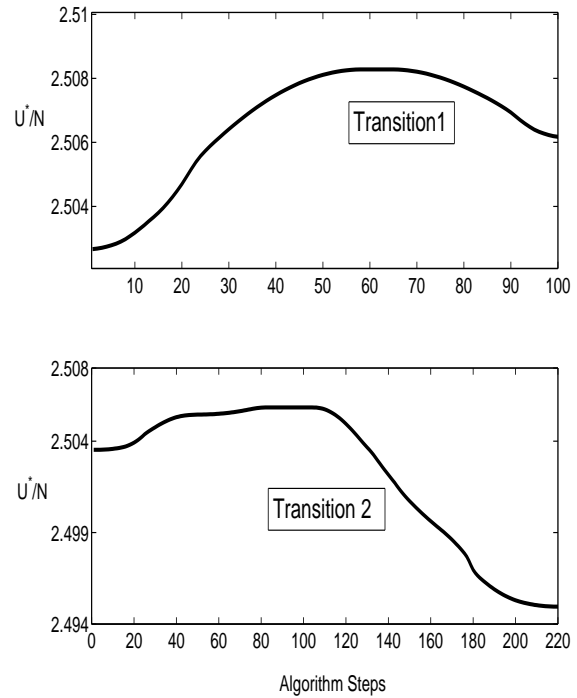} 
\caption{``Reaction coordinate" for the two marked transitions in Fig. \ref{energy}. These graphs are obtained by starting at the saddle point and moving in the direction of the (unique) unstable eigenvector. Note that in general saddles are not guaranteed to be of order 1, and may have more than one unstable direction.}\label{trans}
\end{figure}

The same method showed similar results for the first transition marked in Fig. \ref{energy}, i.e. a transition between two minima, with the transition state being a \textit{first order} saddle point. It is therefore of interest to find the 
`reaction coordinate' for the transition, or the path on the potential surface which underlies the ageing process. It is very difficult to do so using the molecular dynamics simulation, and to achieve this
we used the method of `eigenvector following' (see appendix for details). Starting from the saddle, one takes a small step in one of the directions of the negative eigenvalue and then uses the eigenvector following algorithm to trace the path leading to the minimum lying in this direction. This method works very well, resulting in the reaction coordinates shown in Fig. \ref{trans}. It is important to notice that the energies of the minima
are indeed in agreement with the results of the previous algorithm. We thus believe that the result shown
connect indeed between the picture in time and the picture in energy landscape, and the reaction coordinate
is the one corresponding to the temporal events shown in Fig. \ref{energy}

\section{Other models}
\label{models}
The emerging insight from the analysis presented in the pervious section is that
the reason for the slowing down in glass forming systems is the generation of clusters
of ground and close to ground states. These clusters have relatively high local shear moduli,
since they are mechanically stable and can support locally a significant amount of shear without
flowing. In between the clusters is where the ageing events take place, and indeed in those regions
the local shear moduli are small, even negative, depending on the resolution of the calculation of the local shear modulus. When the temperature gets lower, the clusters increase in size, but there is no point of crisis where the relaxation time diverges unless $T=0$ or for some reason the system crystalizes. Note that the lack of such finite temperature crisis was shown rigorously for this particular model of binary mixture in \cite{08EP}. One question that we need to ask is whether this behavior is generic to glass formers or is it a peculiarity of the present model of binary mixture.

In the context of our own work we have considered recently two other, very different, models of
glass formers, i.e. the Shintani-Tanaka model \cite{06ST} and dry glycerol \cite{05PHRBFKB}. In both systems it was demonstrated, using quantitative theory, that the scenario discussed here appears general, independent of the
very different details characterizing these systems \cite{07ILLP,08HP}. Thus for example we could show in the context
of the Shintatni-Tanaka model that there exists a typical sale $\xi$ that
dominates the mean relaxation time as a function of the temperature. In other words, the $\tau_\alpha$
relaxation time in this model could be quantitatively determined by a formula reading
\begin{equation}
\tau_\alpha = \tau_0 \exp(A \xi /T) \ ,
\end{equation}
where $A$ is a temperature-independent dimensional constant and $\tau_0$ is the microscopic attempt time. A similar result was demonstrated for the present binary model in \cite{06ABIMPS}. The non-Arrhenius behavior of this formula is due to the strong increase of $\xi$ when
the temperature decreases, due to the increase in the size of the clusters, but without any singularity at any temperature $T>0$. For the glycerol one could go even further, predicting not only the mean relaxation time but also computing the functional form
of the dielectric response function in good relation to experiments. Here one argued that each cluster of $s$ molecules contributes an exponential decay of its dipole moment with an $s$-dependent decay constant. Once averaged over all the clusters, the resulting  response function
is strongly non-exponential as is indeed observed. The main point is that the long relaxation times
are again due to the large clusters (smaller cluster may contribute $\beta$ peaks when the
conditions are right, cf. \cite{08HP}). As long as the system does not crystalize there is no mechanism
for a crisis of the type predicted by the Vogel-Fulcher formula or the Kauzmann argument. For more
detailed discussion of this important issue the reader is referred to \cite{08EP}.

\section{Summary and Discussion}
\label{summary}

In summary, we focused on generic ageing events in our glassy system, to understand the relation
between slowing down and spatial inhomogeneity. We have shown that the inhomogeneity can
be seen in two complimentary ways, one by observing the clusters of competing ground and close to ground states, and the other by measuring the local shear modulus. The latter quantity fluctuates from place to place, tending to be high and positive in the presence of clusters and low and even
negative outside the clusters. Ageing events occur in these mechanically unstable regions, and they
result in a local re-organization, usually increasing the area spanned by clusters. We examined
closely the nature of these ageing events, and showed again that they can be described in two complementary forms. One is a localized chain of movements, a so-called ``excitation chain" (accompanied by a larger reversible collective motion) and the second is a transition between two inherent states, or local minima in the potential surface of the system, with the transition crossing a saddle point of order unity. Note that we could differentiate between the very localized permanent change and the ``string-like" event that is in part reversible.

We proposed, on the basis of the present model of a binary mixture and other systems that were analyzed recently in the same spirit, that the results obtained here are generic, and that there exists
a wide class of glass forming systems whose theoretical understanding calls for an analysis
of its distribution of clusters of competing ground and close to ground states; in a future publication
we will show that the temporal properties of such systems can be completely understood in these
terms.

\appendix
\section{The ground state}

The average potential energy per particle  in a binary mixture with the 
interactions defined by the potential (\ref{e1}) is given by:
\beq
\frac{U^{*}}{N}=\frac{1}{2\cdot N}\sum\limits_{i\ne j}\left(
\frac{\sigma_{ab}}{r_{ij}}\right)^{12},
\label{engen}
\eeq
where $r_{ij}$ is the distance between particles $i$ and $j$.

For the equimolar system of $N\to\infty$ separated small and large particles 
(see the upper panel of Fig.~\ref{qqc}) the interactions between the two
species is neglible ($o(1/\sqrt{N})$) and the total energy  
(\ref{engen}) can be written as:
\beq
\frac{U^{*}}{N}=\frac{3}{2}\Bigg[\left(\frac{\sigma_{11}}{r_{11}}\right)^{12}+
\left(\frac{\sigma_{22}}{r_{22}}\right)^{12} \Bigg]
\label{enh}
\eeq
where $r_{11}$ and $r_{22}$ are the distances between small and large
particles respectively.

The area of a hexagon is  $\frac{\sqrt{3}}{2} r^2$, where $r$ is 
the distance between particles. The dimensionless total area $A^*=1/\rho ^*$ 
per particle of the 
system of separated small and large particles is given by:
\beq
\frac{A^*}{N}=\frac{\sqrt{3}}{4}\cdot 
(r_{11}^2+r_{22}^2).
\label{ah}
\eeq
The enthalpy depends on two variables $r_{11}$ and $r_{22}$, so the minimum of 
this function is defined by two equations $\partial H^*/ \partial r_{11}=0$ and 
$\partial H^*/ \partial r_{22}=0$. Solutions of these equations and evaluated 
values for the energy, enthalpy and density are displayed in Tab. \ref{tab1}.

\begin{table}[!h]
\centering
\caption{a-NPT ensemble, b-NVT ensemble.}
\begin{tabular}{|c|c|c|c|c|}
\hline
Phase&$Sep.^a$&$Mixt.^a$&$Sep.^b$&$Mixt.^b$ \\
\hline
$r_{11}$&1.031&1.003&1.039&0.995 \\
\hline
$r_{22}$&1.376&1.473&1.386&1.462 \\
\hline
$r_{12}$&$-$&1.187&$-$&1.179 \\
\hline
$P^*$&13.5&13.5&12.2&14.9 \\
\hline
$\rho ^*$&0.781&0.77&0.77&0.781 \\
\hline
$U^*/N$&2.881&2.922&2.648&3.179 \\
\hline
$H^*/N$&20.168&20.453&18.752&22.257 \\
\hline
\end{tabular}
\label{tab1}
\end{table}

In the configuration where small and large particles are mixed (the lower 
panel of Fig.~\ref{qqc}) each small particle interacts with one small particle 
and four large particles. In turn, each large particle interacts with 
three neighboring large particles and four neighboring small particles. Therefore, the general 
expression (\ref{engen}) is simplified to:
\beq
\frac{U^{*}}{N}=\frac{1}{4}\Bigg[\left(\frac{\sigma_{11}}{r_{11}}\right)^{12}+
8 \left( \frac{\sigma_{12}}{r_{12}} \right)^{12}
+3\left(\frac{\sigma_{22}}{r_{22}}\right)^{12}\Bigg],
\label{enm}
\eeq  
where in addition to  (\ref{enh}) the contribution of interactions between
small and large particles separated by $r_{12}$ are taken into account.

Any configuration of homogeneously mixed small and large particles can be 
constructed using 'boat' unit \cite{LSW01} (see the lower panel of 
Fig.~\ref{qqc}) which consists of one 'thin' and three 'fat' rhombi.
There are $1/4 N$ 'thin' rhombi and $3/4 N$ 'fat' rhombi in the system, so 
the area per particle is given by:
\beq
\frac{A^*}{N}=\frac{1}{4}\cdot (A^*_T+3\cdot A^*_F),
\label{ar}
\eeq
where $A^*_T$ is the area of a thin rhombus and $A^*_F$ that of a fat one.

It follows from the 'boat' design that the distances between particles are 
not independent. It is useful to introduce a new variable $\alpha$, the small 
angle of the 'fat' rhomb, so that the large angle of the 'thin' rhomb is 
$2\pi-3\alpha$. All rhombi have sides of the same size $r_{12}$ and  the 
distances between small-small and large-large particles can be expressed as:
\bea
r_{11}&=&-2\cdot r_{12}\cdot cos(\frac{3}{2}\alpha) \nonumber \\
r_{22}&=& 2\cdot r_{12}\cdot sin(\frac{1}{2}\alpha )
\label{da}
\eea

The area per particle of the mixed system (\ref{ar}) as a function of these 
variables is given by:
\bea
\frac{A^*}{N}&=&\frac{1}{4}\cdot r_{12}^2\cdot  
\left(-sin(3\alpha)+3\cdot sin(\alpha)\right) \nonumber \\
&=&r_{12}^2\cdot sin^{3}(\alpha).
\label{arang}
\eea

Substitution of (\ref{da}) into (\ref{enm}) yields the following expression 
for the energy:
\bea
\frac{U^{*}}{N}&=&\frac{1}{4}\left(\frac{1}{2\cdot r_{12}}\right)^{12}\cdot
\Bigg[\left(\frac{\sigma_{11}}{cos(\frac{3}{2}\alpha)}\right)^{12} 
\nonumber \\
&+&8\left(\sigma_{12}\right)^{12}+3 \left(\frac{\sigma_{22}}
{sin(\frac{1}{2}\alpha)}\right)^{12}\Bigg],
\label{enma}
\eea

The minimum of the enthalpy in this case is defined by equations 
$\partial H^* /\partial r_{12}=0$ and $\partial H^* /\partial \alpha =0$. 
The solution for the angle $\alpha$ depends only on the ratio 
$\sigma_{22}/\sigma_{11}$ and for the system under consideration 
$\alpha=76.65^{o}$. The interparticle distances,  energy, enthalpy and 
density are shown in Table~\ref{tab1}.

In the NVT ensemble it is necessary to find the minimum of the energy given 
by (\ref{enh}) or (\ref{enma}) under the constraints (\ref{ah}) or 
(\ref{arang}), respectively. The calculation results are also presented in 
Tab.\ref{tab1}.

\section{Locating extrema}
\label{extrema}
Finding a local minimum of a function of several variables is a common problem in applied math. There are several methods for doing this, some of them like the Newton-Raphson method are quite simple. Finding a saddle point, on the other hand, is a more subtle problem since it is not clear if the appropriate direction to go is up or downhill. In our work we have used two different methods to overcome this problem:

\subsection{Square gradient minimization}

One way of finding saddle points is by finding a local minimum of the squared modulus of the
gradient of the original potential function:
\begin{equation}
W(\mathbf{x})=|\nabla U(\mathbf{x})|^2 = \sum_{i=1}^N \sum_{\alpha=1}^2 \left( \frac{\partial U}{
    \partial x_{i,\alpha} } \right)^2.
\end{equation}
Since $W(\textbf{x})$ is nonnegative, at the absolute minima $W(\textbf{x})=0$, which
implies $\nabla U=0$ (a saddle point). An efficient way to minimize such functions is the Newton-Raphson method.
At each algorithm step $x_{n+1} = x_n + \delta x_n$ the following equation should be satisfied:
\begin{equation}
\nabla U(x_n + \delta x_n) = 0 \ .
\end{equation}
This equation is solved to first order:
\begin{equation}
\nabla U(x_n) + \delta x_n \nabla^2 U(x_n)  = 0 \ .
\end{equation}
Inverting for $\delta x_n$ one finds
\begin{equation}
\delta x_n = -\nabla^2 U(x_n) \nabla U(x_n) = -\hat{H} \nabla U(x_n) \ .
\end{equation}

We have used a publicly available numerical algorithm - the LBFGS algorithm (\cite{liu89, netlib}) which uses an approximate version of the Newton method. 
The biggest drawback in minimizing the function $W(\textbf{x})$ instead of minimizing $U(\textbf{x})$ directly is that the minimization often converges to a local minimum of $W(\textbf{x})$ , which is not a saddle point of $U(\textbf{x})$. Threfore, when minimizing a molecular dynamics realization of a system travelling in the energy landscape, not all of the instantaneous states converge to a minimum or a saddle of $V(\textbf{x})$ and some information is lost.

\subsection{Eigenvector following}

In order to demonstrate the topological connectedness of two minima
separated by a saddle we have used the Eigenvector Following Method 
following the algorithm as described by Wales and coworkers \cite{wales93, wales94, wales96} (the 
original idea for the algorithm was proposed by Cerjan and Miller \cite{cerjan81}).
At each iteration a step $\Delta x$ is proposed,
which in the base that diagonalizes the Hessian (at the specific point in the 
algorithm path) is
\cite{wales93,wales94}
\begin{equation}
  \Delta x_\mu = S_\mu \frac{{2 g_\mu }}{ |h_\mu| \left( 1 + \sqrt{1 + 4 g^2_\mu
        / h^2_\mu } \right)} ,
\end{equation}
where $h_\mu$ are the eigenvalues of the Hessian and $g_\mu$ are the
components of the gradient in the diagonal base ($\Delta x_\mu$ is set
to 0 for the directions where $h_\mu=0$; i.e.\ uniform displacements).
The sign $S_\mu=\pm 1$ is chosen as explained below. Note that as
$g_\mu\to0$,
\begin{equation}
  \Delta x_\mu = - \frac{g_\mu}{h_\mu} + O(g_\mu^2), \qquad g_\mu\to0
\end{equation}
where the first term is the Newton-Raphson step. 

If $S_\mu=-1$ the algrithm converges to a minimum along
$\mu$. Therefore, starting from a saddle of order one, setting all of the $S_\mu$ to $-1$ we were able to 
reach the 2 minima separated by this saddle. 

\end{document}